\documentclass[12pt]{article}
\usepackage{graphicx,amssymb,amsmath,amsthm}
\usepackage{cite}

\newtheorem{proposition}{Proposition}

\newcommand{\Li}{[L_1, L_2]}
\newcommand{\Lp}{[L_1^{'}, L_2^{'}]}
\newcommand{\be}{\begin{equation}}
\newcommand{\ee}{\end{equation}}
\newcommand{\bea}{\begin{align}}
\newcommand{\eea}{\end{align}}
\newcommand{\ben}{\begin{eqnarray}}
\newcommand{\een}{\end{eqnarray}}

\newcommand{\spann}{{\mbox{\rm{span}}}}

\newcommand{\ov}{\overline}
\newcommand{\ra}{\rangle}
\newcommand{\la}{\langle}

\newcommand{\sumn}{\sum_{n=1}^N}

\newcommand{\spa}{,\,}

\newcommand{\subz}{n \in \N}
\newcommand{\sumz}{\sum_{\subz}}
\def\N{\mathbb{N}}
\newcommand{\h}{{\cal{H}}}
\newcommand{\W}{{\cal{W}}}

\title{On the truncation of the harmonic oscillator wavepacket } 

\author{L. Rebollo-Neira and S. Jain\\
Aston University,\\
Birmingham B4 7ET, United Kingdom}

\date{}
\begin{document}
\maketitle
\baselineskip = 1.7 \baselineskip

\begin{abstract}
We present an interesting result regarding the 
implication of truncating the wavepacket of the 
harmonic oscillator. We show that
 disregarding the  non-significant tails of 
a function which is the superposition of 
eigenfunctions of the harmonic oscillator 
has a remarkable consequence. Namely, there exit 
infinitely many different superpositions giving rise to 
 the same function on the interval. Uniqueness, in the 
case of a wavepacket, is restored by a postulate of  
quantum mechanics. 

\end{abstract}
\section{Introduction}
We analyse the effect of truncating the wavepacket of the 
harmonic oscillator in the light of the {\em frame theory}. 
Such a theory, developed in 1952 by  Duffin and Shaffer 
in the context of harmonic analysis \cite{da} has been 
applied, for over fifteen years, to construct 
coherent states. More specifically, we 
should mention affine coherent states, 
also called wavelets, 
and  Weyl-Heisenberg coherent states, 
also  known as Gabor frames 
\cite{do,wa,ka1,ali3,ali1,do2,han,alib,rewf,rs}.
We recall in the 
next paragraph the general definition of frames and a
few properties, which is all what we need for the 
purpose of the present effort. For a complete treatment 
of frames we refer to \cite{yo,ftf,c2}.  

Given a Hilbert space $\h$, a family 
$\{\phi_n\}_{n\in \N}$ in $\h$ 
is called a {\em{frame}} for $\h$ if 
for every $f \in \h$ 
there exists a pair of constants
$0 < A \le B <\infty $ such that
\be
A \la  f , f \ra  \le  \sum_{\subz} | \la \phi_n , f \ra|^2 \le
B \la  f , f \ra.
\label{fc}
\ee
The constants $A$ and $B$ are called the
{\em {frame bounds}} and (\ref{fc}) the 
{\em frame condition}. 
From its definition it is clear that a frame
is a complete set,  since
the relations $ \la \phi_n, f\ra=0 \spa  \subz$ imply
that $f\equiv 0$. The removal of an element
from a frame leaves either a frame or an incomplete set.
A frame that ceases to be complete if
an arbitrary element $\phi_n$ is removed, is called
{\em{exact}}. This last property implies that only 
exact frames are bases;
in the general case
the  family $\{\phi_n\}_{n\in \N}$ may be
 over complete.
When the condition $A=B$ holds, the frame is said to be
a {\em{tight}} frame. Assuming that all the elements 
$\phi_n,\,  \subz $ are normalised to unity,
a tight frame is an orthogonal basis if and only if $A=B=1$.
Notice that this implies that a tight frame with
frame bounds $A=B=1$ is an orthogonal basis only if
the elements are normalised to unity, otherwise it
is a redundant frame. 

In this paper we introduce a class of redundant tight frames 
which are trivially obtained 
by redefining the functions of an orthonormal basis  
to be zero outside an interval. Such a restriction allows 
us to use the truncated functions to represent a given function  
vanishing outside the identical interval. This has 
a remarkable consequence, namely, the 
coefficients of the corresponding linear span are {\em{not unique}}. 
We discuss this important consequence in relation to the truncation of 
the wavepacket of an harmonic oscillator. We show that, 
by disregarding the
non-significant tails of a function which is  the
superposition of the harmonic oscillator eigenfunctions 
one creates a {\em{null space}}. As a consequence, the 
restriction of the wavepacket to the reduced interval can be 
realized by infinitely many different coefficients 
giving rise to the same function on the interval. This 
is, certainly, a striking result. 
Nevertheless, uniqueness can be 
restored by means of one of the postulates of 
 quantum mechanics. 

The paper is organised as follows: In Section II we 
 introduce the construction of tight frames for 
the Hilbert space of functions vanishing 
outside an interval 
by simple truncation of orthonormal basis functions. 
In Section III we apply these results to analyse the effect of 
truncating the wavepacket of the harmonic oscillator. The 
conclusions are drawn in Section IV. 

\section{Building tight frames from orthonormal bases}
The next proposition shows that, for the space of
square integrable functions vanishing outside an interval, 
one can construct tight frames by simple restriction of 
orthonormal functions to the corresponding interval. 
\begin{proposition}
\label{p1}
Let $\{\psi_n\}_{\subz}$ be an orthonormal basis for $L^2 \Li $, i.e.,
\be
\la \psi_m, \psi_n \ra =\int_{L_1}^{L_2} \psi_m^\ast(x)\psi_n(x)\, dx =
\delta_{m,n}
\ee 
and ${\chi}_{\Lp}$ the characteristic function for the 
interval ${\Lp}$, i.e, 
$${\chi}_{\Lp}(x)= \left\{\begin{array}{cll} 1
 \, \, & {\mbox{if}} \,\, x \in \Lp \\
                       0 \,\,& {\mbox{otherwise.}}
\end{array} \right.$$ 
Functions $\{\psi'_n\}_{\subz}$, obtained as
$\psi'_n(x) ={\chi}_{\Lp}(x)\psi_n(x) \spa \subz$, with
 $\Lp \subset \Li$, constitute a tight frame for the
subspace ${\cal{W}}$ of square integrable functions
vanishing outside $\Lp$.
\end{proposition}
\begin{proof}
On the one hand, since $\{\psi_n\}_{\subz}$ is an
orthonormal basis of $L^2\Li$, for all $f \in {L^2\Li}$ we have
\be
||f||^2= \la f,f \ra=
\sumz \la f , \psi_n \ra \la  \psi_n,  f \ra.
\ee
On the other hand, for all $f \in {\cal{W}}$ it is true
that ${\chi}_{\Lp} f= f$ and we further have:
$$\sumz \la f , \psi_n' \ra \la  \psi_n', f \ra =
 \sumz \la f {\chi}_{\Lp}, \psi_n \ra \la \psi_n,
{\chi}_{\Lp} f \ra
  = ||f||^2,$$
which proves that $\{\psi_n\}_{\subz}$ is a redundant tight
frame for ${\cal{W}}$, since the elements $\psi_n'$ are not normalised 
to unity.
\end{proof}
We prove next that, through the finite subset of 
frames elements $\psi_1',\ldots,\psi_N'$
constructed as indicated in Proposition~\ref{p1} 
we can construct the orthogonal 
projection of $f$ onto 
${\cal{S}}=\spann\{\psi_1,\ldots,\psi_N\}$, restricted to 
$\Lp$.
\begin{proposition}
Let $\psi_1',\ldots,\psi_N'$ be as defined in 
Proposition~\ref{p1}, and for each $f \in {\cal{W}}$ let us define  
a function $f^N$ as:
\be
\label{opp}
f^N= \sum_{i=1}^N \psi_i' \la \psi'_i, f \ra.
\ee
The function given in \eqref{opp} satisfies: 
$$f^N= \left\{\begin{array}{cll} \hat{P}_{\cal{S}}f(x)
 \, \, & {\mbox{if}} \,\, x \in \Lp \\
                   0 \,\,& {\mbox{otherwise,}}
\end{array} \right.$$
where $\hat{P}_{\cal{S}}$ stands for the orthogonal projector 
 operator onto ${\cal{S}}$.
\end{proposition}
\begin{proof}
Since $\{\psi_1,\ldots,\psi_N\}$ is an orthonormal set for
$L^2 \Li$, the orthogonal projection of $f \in L^2 \Li$  onto 
${\cal{S}}$ is given as
\be
\hat{P}_{\cal{S}}f = \sum_{i=1}^N \psi_i \la \psi_i, f\ra. 
\ee
For $f \in {\cal{W}}$ it holds  that
$$\sum_{i=1}^N \psi_i \la \psi_i, f\ra= \sum_{i=1}^N \psi_i \la \psi_i', f\ra.$$Then 
\be
\hat{P}_{\cal{S}}f = \sum_{i=1}^N \psi_i \la \psi_i', f\ra,
\ee
and the proof follows by multiplication of both sides of the equation 
by ${\chi}_{\Lp}$. 
\end{proof}
A convenient property of tight frames is that 
 the coefficients of a linear span in a tight frame superposition
are obtained by inner products with the frame functions.
In our context this implies that  
the coefficients of an orthonormal 
expansion and the ones to span a function
in $\W$ by means of the frame in Proposition~\ref{p1}
 are computed in an equivalent manner.
The essential difference is that, as discussed below,
the coefficients in the frame superposition are 
{\em{not unique}}.
 
Let us consider the frame constructed in Proposition \ref{p1} and 
let us define ${\cal{S}'}=\spann\{\psi'_1,\ldots,\psi'_N\}$. For every 
$f\in {\cal{S}'} \subset {\cal{W}}$ we have: 
\be
\label{fd}
f=\sum_{n=1}^{N} 
c_n \psi'_n, \,\,\,\,\,\text{with}\,\,\,\,\,c_n=\la \psi_n', f \ra.
\ee
Since $f \in \W$ it is true that $c_n=\la \psi_n', f \ra= \la \psi_n, f\ra$ 
and the equivalence with the orthonormal case follows. However, the 
redundancy of the frame implies that the coefficients in \eqref{fd} are 
not unique. Indeed, since a redundant frame is linearly dependent,  
the following situation can occur:
\be
\label{ld}
0=\sum_{n=1}^{N} c'_n \psi'_n,\,\,\,\,\,\text{for}\,\,\,\,\,\sum_{n=1}^N |c'_n|^2 \ne 0.
\ee
Taking inner product both sides of the equation on the left 
with each 
 $\psi'_m$   we have
\be 
0=\sum_{n=1}^{N} c'_n \la \psi'_m , \psi'_n \ra,
\ee
that we can recast as 
\be
\label{ld2}
0= G \vec{c'}
\ee
where $G$ is a matrix of elements 
$g_{m,n}= \la \psi'_m , \psi'_n \ra \,,n,m= 1,\ldots,N$ and
$\vec{c'}$ a vector, the component of which are 
the coefficients $c'_n,\, n= 1,\ldots,N$. 
Equations \eqref{ld} and \eqref{ld2} imply that the 
general form for \eqref{fd}  is
\be
\label{eq}
f=\sumn c_n \psi'_n + \sumn c'_n \psi'_n,
\ee
with $c_n, \,n=1,\ldots,N$ given in \eqref{fd} and 
$c'_n,\, n=1,\ldots,N$ the components of a vector 
$\vec{c'} \in$ null($G$). 

It is appropriate to stress the significance of 
Proposition~\ref{p1} when the interval $\Li$ is 
actually the whole real line. Then, for numerical 
calculations one is obliged 
to work on a finite domain. An important consequence of this 
fact will be discussed in the next section.

\section{On the truncation of the harmonic Oscillator 
wavepacket}
We show here that the results of the previous section 
are relevant to the analysis of the truncation of the
wavepacket of the harmonic oscillator. To this end 
we simulate two different situations, which are specially 
devised to illustrate the phenomenon we wish to 
discuss. 

\begin{figure}[h!]
\begin{center}
\mbox{\includegraphics[angle=0,width=6.5cm]{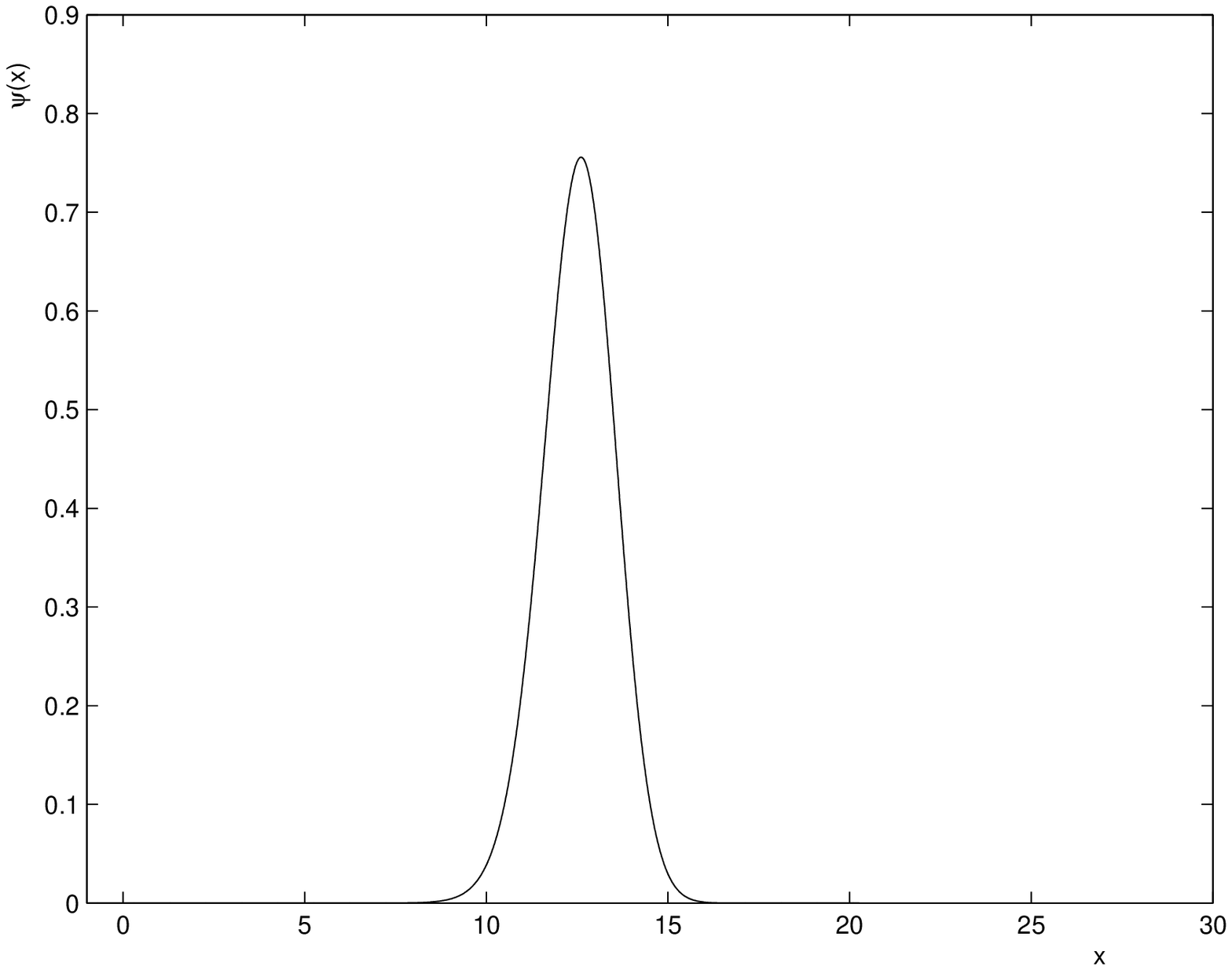}
\includegraphics[angle=0,width=6.5cm]{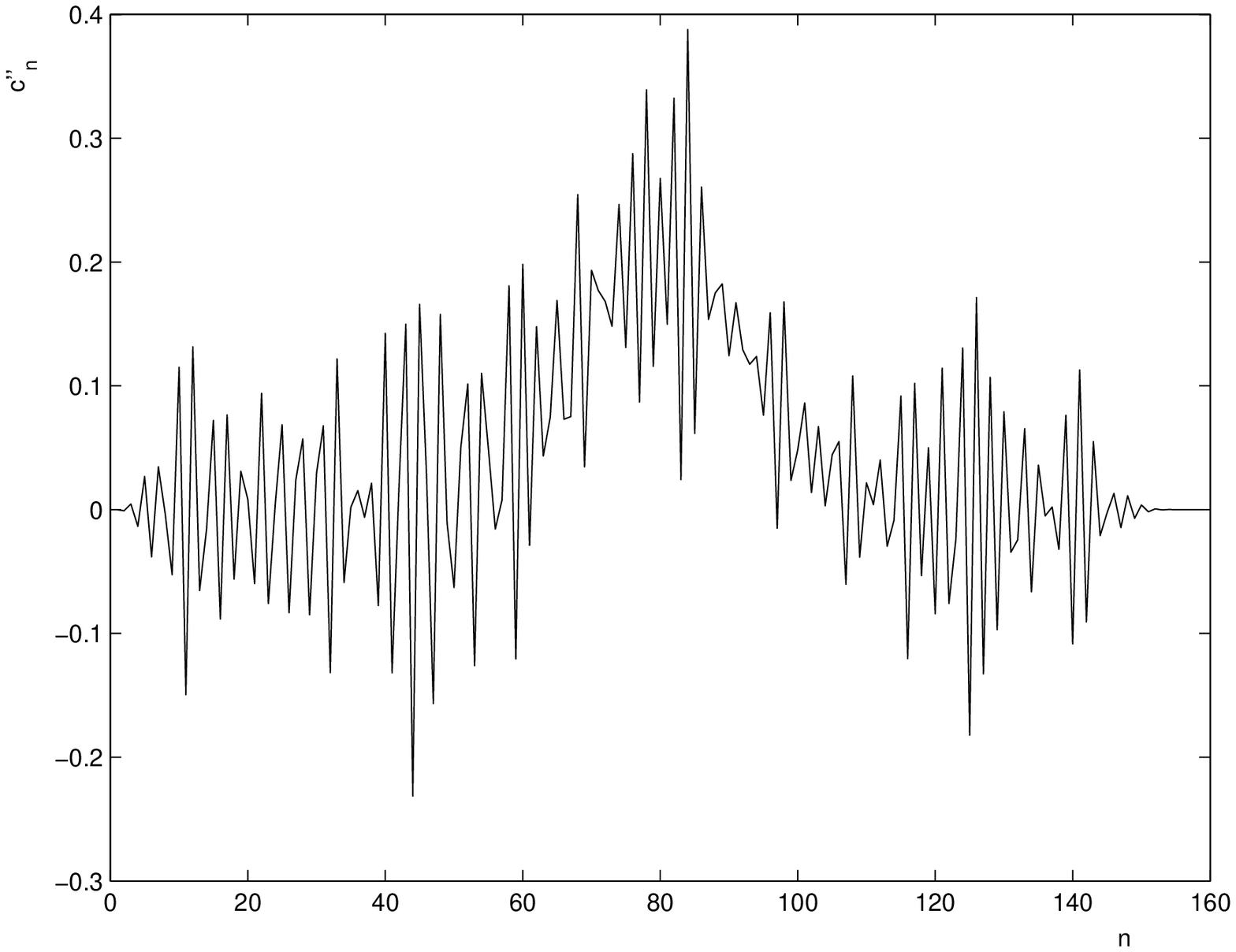}}
\caption{\small{The graph on the left 
represents the function $\Psi$ 
constructed by using the coefficients given in \eqref{cn} 
and also the coefficients $c''_n$ plotted
in the right graph.}}
\label{wp1}
\end{center}
\end{figure}

Let us consider that a normalised to unity function 
$\Psi(x) \in \h$ is generated as linear 
superposition of the harmonic oscillator eigenfunctions, 
i.e., 
\be
\Psi(x)=
\sum_{n=1}^{N} c_n\frac{e^{-0.5x^2} H_n(x)}
{\sqrt{2^{n-1}(n-1)!\sqrt{\pi}}}, \,\,\,\,\,\,n,m=1,\ldots,N,
\label{wp}
\ee
where we have written 
$\psi_n(x) =\frac{e^{-0.5x^2}H_n(x)}{\sqrt{2^{n-1}(n-1)!\sqrt{\pi}}}$ in terms of the 
 Hermite polynomial $H_n(x)$ . The 
coefficients $c_n,\, n=1,\ldots,N$ in (\ref{wp}) 
are simulated according to the equation:
\be
\label{cn}
c_n=\frac{e^{-0.0032(n-80)^2}}{\sqrt{\sum_{n=1}^N e^{-0.0064(n-80)^2}}}.  
\ee
We consider $N=160$. 
The left graph of Figure~\ref{wp1} depicts 
the corresponding function for $x\in[-1,30]$. 
Since the values $\Psi(x)$ are 
very small outside this interval ($\Psi(-1)=8.7 \times 10^{-11}$ 
and $\Psi(30)= 2.7 \times 10^{-92}$) one does not commit a significant 
error by calculating expectation values using this domain. In order to 
illustrate this fact let us calculate $\ov{x}$ and $\ov{x^2}$ 
as
\ben
\ov{x}&=& \int_{-40}^{40} |\Psi(x)|^2 x\,dx = 12.56967570231863 
\nonumber \\
\ov{x}&= &\int_{-1}^{30} |\Psi(x)|^2 x\,dx = 12.56967570231863
\een
\ben
\ov{x^2}&= &\int_{-40}^{40} |\Psi(x)|^2 x^2\,dx = 158.4912423027778
\nonumber \\
\ov{x^2}&= &\int_{-1}^{30} |\Psi(x)|^2 x^2\,dx = 158.4912423027777.
\een
The difference in the values of  $\ov{x}$ 
cannot be observed in the given format and the values of $\ov{x^2}$
differ only in the last of the 16 digits.
Thus one could conclude that, 
for the purpose of computing expectation values, neglecting the   
tails of the distribution outside the interval $x\in[-1,30]$  is not harmful. 
However, the assumption that $|\Psi(x)|=0$ for 
$x \notin [-1,30]$ has a tremendous consequence: 
the coefficients of the  superposition (\ref{wp}) 
are thereby not unique.
In order to illustrate this we compute the vectors 
in the null space of 
matrix $G$ of elements 
\be
g_{m,n}= \int_{-1}^{30} H_m(x)H_n(x)\frac{e^{-x^2}}
{\sqrt{2^{m-1}(m-1)! 2^{n-1}(n-1)!}\pi}\,dx.
\ee
We use just one of the  eigenvectors spanning null$(G)$, 
say the vector $\vec{c'}$,  to construct the coefficients 
$c''_n=c_n + c'_n, \,n=1,\ldots,160$, with $c_n$ as in \eqref{cn} 
and $c'_n, n=1,\ldots,160$ the components of $\vec{c'} \in$ null$(G)$.
The right graph in Figure~\ref{wp1} plots the coefficients $c''_n$. 
We use now these coefficients to construct the function $\Psi''(x)$.  
By calculating 
\be
||\Psi'' - \Psi|| =\sqrt{\int_{-1}^{30}|\Psi''(x) - \Psi(x)|^2\,dx}=
 1.4198\times 10^{-18}
\ee
we do not see any significant difference in the functions. 
However $||\vec{c''}-\vec{c}||=||\vec{c'}||=1$, which clearly shows that 
the  function are considering can be generated in
many different ways on the interval $[-1,30]$. Even considering the 
interval $[-10,40]$ null($G$) is still not empty. It would be empty for 
$[-15,40]$, though, but only  if the maximum 
number of states $N=160$ is maintained fixed.  

It should be noted that, although the `critical' interval 
depends on the number $N$ and a smaller value of $N$ would 
decrease the length of the critical interval, 
there is room for adjustments. 
Indeed, by considering zero the coefficients which do not 
intervene in the original superposition,
but allowing the transformation to have larger dimension, the critical interval is enlarged. 
Notice that this opens the possibility of producing the 
identical function by means of states that were not 
 present in the original superposition. The next example illustrates 
this situation. 

\begin{figure}[h!]
\begin{center}
\mbox{\includegraphics[angle=0,width=6.5cm]{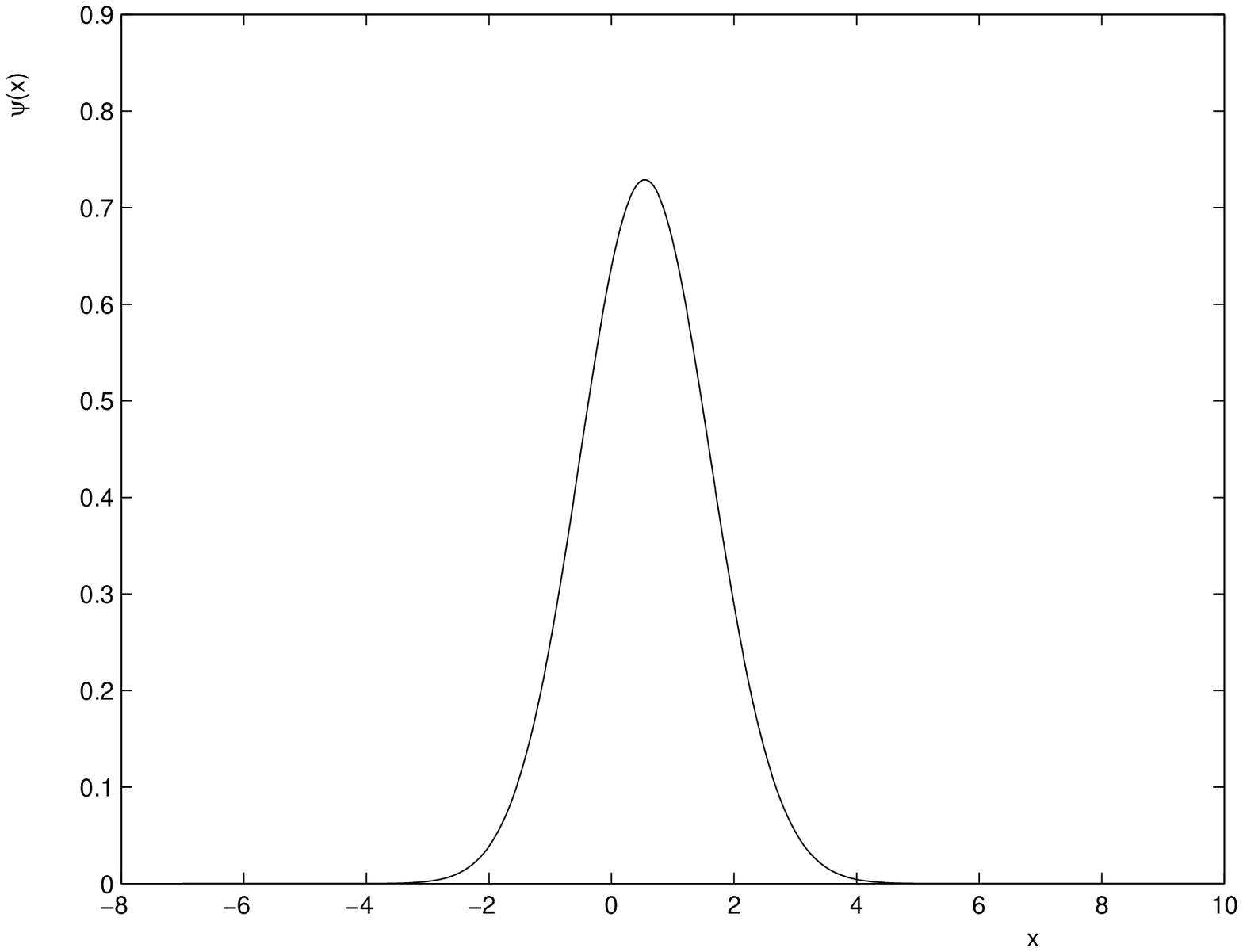}
\includegraphics[angle=0,width=6.5cm]{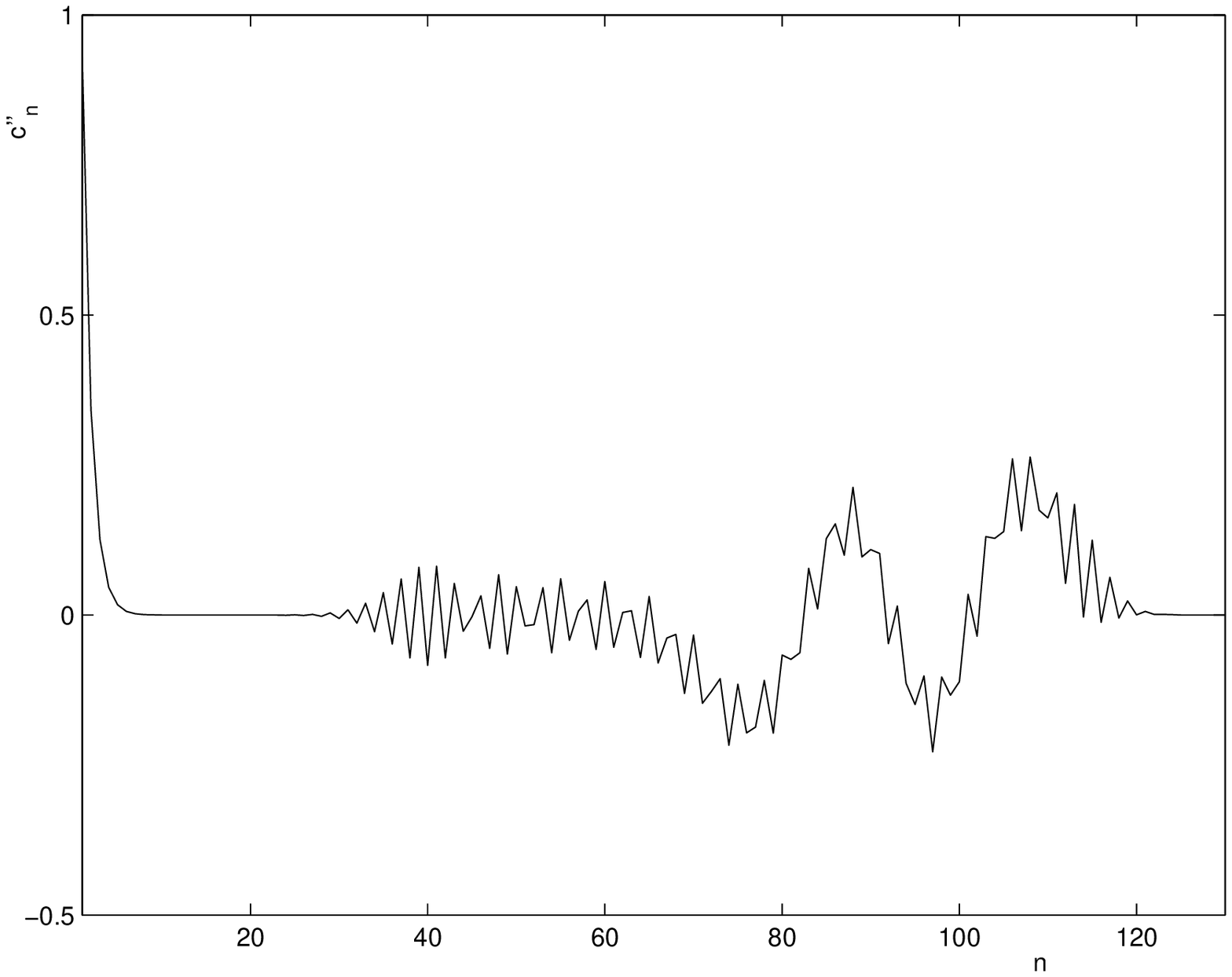}}
\caption{\small{The graph on the left represents the function $\Psi$
constructed by using the coefficients given in \eqref{cn2}
and also the coefficients $c''_n$ plotted in the right graph.}}
\label{wp2}
\end{center}
\end{figure}

Consider that the coefficients of the  superposition 
(\ref{wp}) are now  simulated as 
\be
\label{cn2}
c_n=\frac{e^{-n}}{\sqrt{\sum_{n=1}^{20}e^{-2n}}}, \,\,\,\,\,\,\, n=1,\ldots,20
\ee
The corresponding function $\Psi(x)$ is plotted in the left graph of 
 Figure~\ref{wp2}. In this case the 
critical interval yielding lack of uniqueness is included in the 
main support of $\Psi(x)$. Hence, the restriction to such an interval 
is not possible. Now, increasing the value of $N$ to 130 for instance, 
and considering 
$c_n=0,\,n=31,\ldots,130$, the  function $\Psi(x)$ does not change. 
Nevertheless,  
we have a $130 \times 130$ matrix  $G$  constructed by  
extending the interval to one containing the most significant support of 
 $\Psi(x)$ (in this case $[-7\,,\,10]$). 
Taking one of the vectors in null$(G)$ we
construct the coefficients $\vec{c''}$ 
by an equivalent process as in the 
previous example. With this coefficients, 
plotted in  the right 
graph of Figure~\ref{wp2}, we 
construct the  graph on the left of Figure~\ref{wp2}. 
Notice that, the fact that coefficients $c''_n$ have significant values
for $n=31,\ldots,130$, implies that  $\Psi(x)$  
can be realized by states which were not  present in the  
original superposition. Nevertheless, the average energy is the same. 
Indeed,  
\be
\ov{E}= \la \psi'', \hat{H} \psi'' \ra=\sum_{n=1}^{N} \sum_{m=1}^{N} 
(c_n^{\ast} +{c'_n}^{\ast})\la \psi'_n, \hat{H} (c_m +c'_m) \psi'_m \ra
\ee
and, since by hypothesis $\sum_{m=1}^{N} c'_m \psi'_m,=0$, 
it follows that
\be
\ov{E}= \sum_{n=1}^{N} \sum_{m=1}^{N} c_n^{\ast} \la \psi'_n, \hat{H}
c_m \psi'_m \ra =  \la \psi',  \hat{H} \psi' \ra.
\ee
This result stresses the point that was 
 made initially. Namely, that without normalising the 
coefficients, {\em there  could be infinitely  many 
different ways of realizing the harmonic
oscillator wavepacket on a finite interval}. 
However, according to quantum mechanics
 each $|c''_n|^2$ represents the probability of 
finding the harmonic oscillator in the state $n$. Hence, to 
 be able to maintain this interpretation we must 
 impose the normalisation 
condition on the coefficients $c''_n$. As a consequence, we 
cannot use a vector $\vec{c'}$ of arbitrary norm. In fact, 
in order to construct a normalised vector 
 we need to consider $\vec{c''}= \vec{c}+D \vec{c'}$,
 with $\vec{c'}\in$ null$(G)$
and $D$ a constant to be determined by the condition 
$||\vec{c''}||^2=1$. Writing this condition explicitly 
we have
\be
||\vec{c''}||^2= ||\vec{c}||^2  + D\la \vec{c},\vec{c'} \ra + 
D^\ast\la \vec{c'},\vec{c} \ra 
+|D|^2||\vec{c'}||^2 =1,
\ee
and, since $\vec{c'}$ and $\vec{c}$ are orthogonal to
each other we further have
\be
||\vec{c''}||^2= ||\vec{c}||^2 + |D|^2||\vec{c'}||^2=1.
\ee
The value of $||\vec{c}||^2$ is fixed by the function 
$\Psi(x)$, in our case $||\vec{c}||^2=1$. 
Thus, the only possible solution to the above
equation is $D=0$.  This leads to  the conclusion that 
 amongst all the possible superposition 
giving rise to the  same 
function $\Psi(x)$ on the interval,
there is only {\em one} which is  consistent
with the physical 
significance of quantum mechanics. 

\section{Conclusions}
An interesting result, arising by limiting the domain of a
normalised function which is the superposition
of the harmonic oscillator eigenfunctions,
has been discussed. It was shown that,
when restricted to a finite interval,
such a function can be realized in many different ways.
Although the mean values of the
physical quantities are not affected in any significant
manner,  they can be the result of infinitely many
different combinations of eigenfunctions. Uniqueness is
restored by endowing the function with the significance of a
wavepacket. It was proved that, in that case, 
there is only {\em one} set of coefficients that can 
fulfil the normalisation condition.

\end{document}